\documentclass[12pt]{article}
\usepackage{amsmath,amssymb,latexsym,epsf,epsfig,graphicx,tikz}

\begin{document}

\newcommand{\sect}[1]{\setcounter{equation}{0}\section{#1}}
\renewcommand{\theequation}{\thesection.\arabic{equation}}
\newcommand{\prt}{\partial}
\newcommand{\II}{\mbox{${\mathbb I}$}}
\newcommand{\CC}{\mbox{${\mathbb C}$}}
\newcommand{\RR}{\mbox{${\mathbb R}$}}
\newcommand{\QQ}{\mbox{${\mathbb Q}$}}
\newcommand{\ZZ}{\mbox{${\mathbb Z}$}}
\newcommand{\NN}{\mbox{${\mathbb N}$}}
\def\G{\mathbb G}
\def\UU{\mathbb U}
\def\S{\mathbb S}
\def\T{\mathbb T}
\def\tS{\widetilde{\mathbb S}}
\def\V{\mathbb V}
\def\tV{\widetilde{\mathbb V}}
\newcommand{\D}{{\cal D}}
\def\hint{H_{\rm int}}
\def\R{{\cal R}}

\newcommand{\rd}{{\rm d}}
\newcommand{\diag}{{\rm diag}}
\newcommand{\U}{{\cal U}}
\newcommand{\K}{{\mathcal K}}
\newcommand{\cP}{{\cal P}}
\newcommand{\dQ}{{\dot Q}}
\newcommand{\dS}{{\dot S}}

\newcommand{\pnf}{P^N_{\rm f}}
\newcommand{\pnb}{P^N_{\rm b}}
\newcommand{\hnf}{P^Q_{\rm f}}
\newcommand{\hnb}{P^Q_{\rm b}}

\newcommand{\ph}{\varphi}
\newcommand{\phd}{\widetilde{\varphi}} 
\newcommand{\phs}{\varphi^{(s)}}
\newcommand{\phb}{\varphi^{(b)}}
\newcommand{\phds}{\widetilde{\varphi}^{(s)}}
\newcommand{\phdb}{\widetilde{\varphi}^{(b)}}
\newcommand{\lambdad}{\widetilde{\lambda}}
\newcommand{\tx}{\widetilde{x}} 
\newcommand{\etat}{\widetilde{\eta}}
\newcommand{\phl}{\varphi_{i,L}}
\newcommand{\phr}{\varphi_{i,R}}
\newcommand{\phz}{\varphi_{i,Z}}
\newcommand{\mum}{\mu_{{}_-}}
\newcommand{\mup}{\mu_{{}_+}}
\newcommand{\mupm}{\mu_{{}_\pm}}
\newcommand{\muv}{\mu_{{}_V}}
\newcommand{\mua}{\mu_{{}_A}}
\newcommand{\mut}{\widetilde{\mu}}

\def\a{\alpha}
 
\def\A{\mathcal A} 
\def\H{\mathcal H} 
\def\U{\mathcal U} 
\def\E{\mathcal E} 
\def\C{\mathcal C} 
\def\L{\mathcal L} 
\def\O{\mathcal O}
\def\I{\mathcal I}
\def\der{\partial }
\def\mis{{\frac{\rd k}{2\pi} }}
\def\ri{{\rm i}}
\def\xt{{\widetilde x}}
\def\ft{{\widetilde f}}
\def\gt{{\widetilde g}}
\def\qt{{\widetilde q}}
\def\tt{{\widetilde t}}
\def\tmu{{\widetilde \mu}}
\def\prt{{\partial}}
\def\tr{{\rm Tr}}
\def\inc{{\rm in}}
\def\out{{\rm out}}
\def\li{{\rm Li}}
\def\e{{\rm e}}
\def\eps{\varepsilon}
\def\k{\kappa}
\def\v{{\bf v}}
\def\ebf{{\bf e}}
\def\abf{{\bf A}}
\def\fa{{\mathfrak a}} 


\newcommand{\finprf}{\null \hfill {\rule{5pt}{5pt}}\\[2.1ex]\indent}

\pagestyle{empty}
\rightline{February 2015}

\bigskip 

\begin{center}
{\Large\bf Non-linear Quantum Noise Effects in\\Scale Invariant Junctions}
\\[2.1em]

\bigskip

{\large Mihail Mintchev}\\ 
\medskip 
{\it  
Istituto Nazionale di Fisica Nucleare and Dipartimento di Fisica, Universit\`a di
Pisa, Largo Pontecorvo 3, 56127 Pisa, Italy}
\bigskip 

{\large Luca Santoni}\\ 
\medskip 
{\it  
Scuola Normale Superiore and Istituto Nazionale di Fisica Nucleare, Piazza dei Cavalieri 7, 56126 Pisa, Italy}
\bigskip 

{\large Paul Sorba}\\ 
\medskip 
{\it  
LAPTh, Laboratoire d'Annecy-le-Vieux de Physique Th\'eorique, 
CNRS, Universit\'e de Savoie,   
BP 110, 74941 Annecy-le-Vieux Cedex, France}
\bigskip 
\bigskip 
\bigskip 

\end{center}
\begin{abstract} 
\bigskip 

We study non-equilibrium steady state transport in scale invariant quantum junctions with focus on the particle 
and heat fluctuations captured by the two-point current correlation functions. We show that the non-linear 
behavior of the particle current affects both the particle and heat noise. The existence of domains of enhancement 
and reduction of the noise power with respect to the linear regime are observed. The impact of the statistics is explored. 
We demonstrate that in the scale invariant case the bosonic particle noise exceeds the fermionic one in the common domain of 
heat bath parameters. Multi-lead configurations are also investigated and the effect of probe terminals 
on the noise is discussed.

\end{abstract}
\bigskip 
\medskip 
\bigskip 

\vfill
\rightline{LAPTH-008/15}
\rightline{IFUP-TH 2/2015}
\newpage
\pagestyle{plain}
\setcounter{page}{1}

\section{Introduction} 
\medskip

The study of particle and heat current fluctuations in quantum systems away from equilibrium attracts much attention 
both from the theoretical and experimental point of view.  Such fluctuations generate noise and therefore spoil 
the signal propagation and detection. It is known \cite{L-98}-\cite{SB-06} however that current fluctuations carry also useful   
information, providing the experimental basis of noise spectroscopy. The recent progress \cite{BDD-05, J-13} 
of measurement techniques indicates the great importance of this spectroscopy for gaining a deeper insight in the mechanism of 
non-equilibrium quantum transport at the microscopic level. The latest confirmation of this fact comes   
from the observation of neutral modes \cite{B-10} and spin noise \cite{K-14} 
in fractional quantum Hall edge states via noise spectroscopy. 
At the theoretical side, the derivation of exact results beyond the linear response approximation is fundamental 
in this context, because the current fluctuations are dominated by non-linear effects.

\begin{figure}[h]
\begin{center}
\begin{picture}(700,40)(80,355) 
\includegraphics[scale=1]{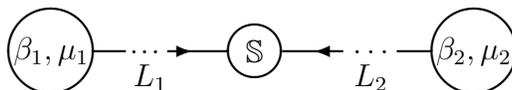}
\end{picture} 
\end{center}
\caption{Two terminal system with scattering matrix $\S$.} 
\label{fig1}
\end{figure} 

The present paper represents a continuation of our study \cite{MSS-14} of non-equilibrium transport
at criticality. The focus here is on the particle and heat current fluctuations of systems with the structure 
shown in Fig. \ref{fig1}. We consider two $(i=1,2)$ semi-infinite leads $L_i$ attached at infinity to the heat reservoirs  
$R_i = (\beta_i,\mu_i)$. The interaction between the leads is localized in the junction and is described by a $2\times 2$ 
unitary scattering matrix $\S$. The system is away from equilibrium if the two leads communicate via 
non-vanishing transmission (off-diagonal) elements of $\S$. It is well known that the physical situation, presented in 
Fig. \ref{fig1}, is nicely described by a Landauer-B\"uttiker (LB) 
non-equilibrium steady state $\Omega_{\beta, \mu}$, constructed from the data 
$(\beta_i,\mu_i)$ and $\S$. We have shown in \cite{MSS-14} that for scale invariant (critical) $\S$-matrices 
the one-point expectation values of the particle and energy currents in the state $\Omega_{\beta, \mu}$ 
can be computed in exact and explicit form, which fully takes into account all non-linear effects and 
reveals the presence of an interesting process of transmutation between heat and chemical potential energy. 
Here we pursue further our analysis, deriving the exact two-point current expectation values, which 
allow to determine the particle and heat current fluctuations in the state $\Omega_{\beta, \mu}$ beyond the linear response 
approximation. In order to test the framework, we first reproduce and generalize some results concerning 
the particle noise. In particular, we show that this noise depends not only on the parameter 
$\mum \sim \mu_1 - \mu_2$, but also on $\mup \sim \mu_1 + \mu_2$. The dependence on 
$\mu_-$ for $\mu_+=0$ and $\beta_1=\beta_2$ has been analyzed in the pioneering works 
of Martin and Landauer \cite{ML-92} and B\"uttiker \cite{B-92} (see also \cite{JB-96,BB-00}). 
A peculiar property of this case is the linear dependence on $\mum$ of 
the particle current flowing in the junction. 
In order to investigate the impact of the non-linear effects in $\mum$, we explore in this paper  
the case $\mup \not=0$ in a systematic way. This regime is of practical interest, because 
the chemical potentials can be varied and measured directly in experiments. 
We show in fact that $\mup$ is the parameter which actually controls the 
noise power in the LB state $\Omega_{\beta, \mu}$. Compared to the case $\mup=0$, we observe a relevant  
noise enhancement for $\mup>0$ and reduction for $\mup<0$. 
An universal upper bound on the particle noise 
away from the scale invariant regime is also established. Following the main steps 
of our particle noise analysis, we derive and study in detail the 
less investigated heat noise as well. 

Both fermions and bosons are 
considered and the influence of the statistics (in particular, of the exclusion principle) is discussed. 
We show that the fermionic noise is a concave function of the transmission probability in the junction, 
whereas the bosonic one is convex. This difference provides a new interesting signature for noise 
spectroscopy. Another remarkable manifestation of the statistics concerns 
the strength of the noise power. We show that for the same heat bath parameters in the sate 
$\Omega_{\beta, \mu}$, the bosonic noise always exceeds the fermionic one.   
A generalization of the framework and results to junctions with $n>2$ leads is also presented. In this case 
we describe the impact of probe terminals on the quantum noise.  

The paper is organized as follows. In section 2 we briefly review the scale invariant Schr\"odinger junction 
and the concepts of particle and heat current fluctuations. In section 3 we compute the particle noise power 
in the fermionic case and describe its general properties, focussing on the non-linear effects for $\mup \not=0$. 
Section 4 concerns the heat noise for fermions. The bosonic particle noise is derived in section 5. 
We also perform there a comparison between the bosonic and fermionic cases. Multi-terminal configurations 
are discussed in section 6. Section 7 is devoted to 
our conclusions. In the appendices A-C we address some technical issues. 

\bigskip 

\section{The scale invariant Schr\"odinger junction} 
\medskip 

Referring for the details to \cite{MSS-14}, we start by recalling those main features of the Schr\"odinger junction 
needed in the derivation of the particle and heat noise below. We consider systems in which the particle number is 
conserved (there is no particle production and annihilation) and denote by $\S^{(n)}$ the scattering 
matrix describing the interaction in the $n$-particle sector. It has been argued in \cite{L-98} that the 
idea of freely propagating particles along the leads accounts remarkably well for the experimental results. 
In fact, the experiments performed with quantum point contacts have shown \cite{K-96} that the simple 
model of independently moving electrons, which allow only for the Pauli principle, fits the 
data with a very good precision. In this spirit we assume that $\S^{(n)}=\II$ for $n>1$ and keep non-trivial 
only the one-body scattering matrix $\S^{(1)} = \S$, which drives the system away from equilibrium 
and is essential in this respect. We concentrate mainly on scale invariant (critical) $\S$-matrices because they are 
simple enough to be analyzed explicitly \cite{MSS-14,Mintchev:2012pe} and are 
expected \cite{BDV-15} to incorporate the basic 
universal features of one-dimensional quantum transport phenomena. 

Let us shortly describe now a concrete realization of this general picture. The dynamics along the leads, 
represented by half lines with coordinates $\{(x,i)\, ,:\, x\leq 0,\, i=1,2\}$ in Fig. \ref{fig1}, is fixed by 
the Schr\"odinger equation\footnote{We adopt the natural units $\hbar =c=k_{\rm B}=1$.}  
\begin{equation}
\left (\ri \prt_t +\frac{1}{2m} \prt_x^2\right )\psi (t,x,i) = 0\, .  
\label{eqm1}
\end{equation} 
The one-body scattering matrix $\S$ is determined by requiring that the bulk Hamiltonian $-\prt_x^2$ admits a 
self-adjoint extension in $x=0$. All such extensions are defined \cite{ks-00}-\cite{k-08} 
by the boundary conditions 
\begin{equation} 
\lim_{x\to 0^-}\sum_{j=1}^2 \left [\lambda (\II-\UU)_{ij} +\ri (\II+\UU)_{ij}\prt_x \right ] \psi (t,x,j) = 0\, , 
\label{bc1} 
\end{equation} 
where $\UU$ is a $2\times 2$ unitary matrix and $\lambda \in \RR$ is a 
parameter with dimension of mass. Eq. (\ref{bc1}) guaranties unitary time evolution and 
defines a specific point-like interaction at the junction. 
The explicit form of the corresponding scattering matrix is \cite{ks-00}-\cite{k-08} 
\begin{equation} 
\S(k) = 
-\frac{[\lambda (\II - \UU) - k(\II+\UU )]}{[\lambda (\II - \UU) + k(\II+\UU)]} \, ,   
\label{S1}
\end{equation} 
$k$ being the particle momentum. 
In order to determine the scale invariant limit of (\ref{S1}) we introduce the 
unitary matrix ${\cal U}$ diagonalizing ${\mathbb U}$, namely 
\begin{equation} 
{\cal U}\, {\mathbb U}\, {\cal U}^* = {\mathbb U}_{\rm d}=  
{\rm diag} \left (e^{-2i\alpha_1}, e^{-2i\alpha_2} \right )\, , 
\qquad -{\pi\over 2} < \alpha_i \leq {\pi\over 2} \, . 
\label{d11}
\end{equation} 
From (\ref{S1}) one can easily deduce 
that ${\cal U}$ diagonalizes also ${\mathbb S}(k)$ for {\it any} $k$. One finds 
\begin{equation} 
{\mathbb S}_{\rm d}(k) = {\cal U}^* {\mathbb S}(k) {\cal U} = \\
{\rm diag} \left ({k+i \eta_1\over 
k-i \eta_1}, {k+i \eta_2\over k-i \eta_2} \right ) \, , 
\label{d3}
\end{equation} 
where 
\begin{equation} 
\eta_i \equiv \lambda {\rm tan} (\alpha_i)\, .  
\label{d4}
\end{equation} 
Scale invariance implies \cite{Calabrese:2011ru} the following alternative  
\begin{equation} 
\eta_i = 
\begin{cases} 
0 \quad \; \;  (\alpha_i=0)\, , & \qquad  \text{Neumann b.c.}\, , \\
\infty \quad (\alpha_i=\pi /2)\, , & \qquad  \text{Dirichlet b.c.} \\ 
\end{cases} 
\label{si1}
\end{equation} 
Accordingly, the set of scale invariant scattering matrices in the family (\ref{S1}) is given by 
\begin{equation}
\S = \U\, \S_{\rm d} \, \U^*\, ,  \qquad \U\in U(2)\, , \qquad \S_{\rm d} = {\rm diag}(1,-1)\,   
\label{sinvS}
\end{equation} 
and the two isolated points $\S=\pm \II$. The latter   
are not interesting because there is no transmission between the two leads and 
the system is therefore in equilibrium. For this reason we concentrate in what follows on (\ref{sinvS}). 
The solution of the problem (\ref{eqm1},\ref{bc1}) is given by \cite{Mintchev:2011mx} 
\begin{equation} 
\psi (t,x,i)  = \sum_{j=1}^2 \int_{0}^{\infty} \frac{dk}{2\pi } \e^{-\ri \omega (k)t}\, 
\left [ \e^{-\ri k x}\, \delta_{ij} +\e^{\ri k x}\, \S_{ij}\right  ] a_j (k) \, , \qquad  \omega(k) = \frac {k^2}{2m} \, .
\label{psi1} 
\end{equation} 
In the fermionic case the operators $\{a_i(k),\, a^*_i(k)\, :\, k\in \RR,\, i=1,2\}$ generate an anticommutation relation algebra 
$\A_+$ defined by $[a_i(k)\, ,\, a_j(p)]_+ = [a^*_i (k)\, ,\, a^*_j (p)]_+ = 0$ and \cite{Bellazzini:2008mn}
\begin{equation}
[a_i(k)\, ,\, a^*_j (p)]_+ = 2\pi [\delta (k-p)\delta_{ij} + \S_{ij}\delta(k+p)] \, , 
\label{rta2}
\end{equation}  
where $*$ stands for Hermitian conjugation. The deformation (\ref{rta2}) of the canonical anticommutation relations 
implements \cite{Liguori:1996xr}-\cite{Mintchev:2003ue} the interaction in the junction. 
The algebra $\A_-$ in the bosonic case is obtained by replacing the anticommutators with commutators. 
The explicit construction of the LB  steady state $\Omega_{\beta, \mu}$ as a linear functional over $\A_\pm$ was given 
in \cite{Mintchev:2011mx}. This state defines the LB representation 
$\H^\pm_{\rm LB}$ of $\A_\pm$, which describes the non-equilibrium 
system in Fig. \ref{fig1} in terms of $(\beta_i,\mu_i)$ and $\S$. For convenience 
we report in appendix A the basic correlation functions of  $\{a_i(k),\, a^*_i(k)\}$ in $\H^\pm_{\rm LB}$. 

In order to compute the particle and heat noise we first recall the form of the particle 
and energy currents 
\begin{equation}
j^N(t,x,i)= \frac{\ri }{2m} \left [ \psi^* (\partial_x\psi ) - 
(\partial_x\psi^*)\psi \right ]  (t,x,i) \, , 
\label{curr1}
\end{equation} 
\begin{equation}
j^E(t,x,i) = \frac{1}{4m} [\left (\partial_t \psi^* \right )\left (\partial_x \psi \right ) 
+ \left (\partial_x \psi^* \right )\left (\partial_t \psi \right ) \\ - 
\left (\partial_t \partial_x \psi^* \right ) \psi - 
\psi^*\left (\partial_t \partial_x \psi \right ) ](t,x,i) \, . 
\label{en1} 
\end{equation} 
The heat current is defined by the combination 
\begin{equation} 
j^Q(t,x,i) = j^E (t,x,i) - \mu_i\,  j^N (t,x,i) \, .  
\label{hc}
\end{equation} 
The quantum fluctuations of any of these currents in the LB state $\Omega_{\beta, \mu}$ is given by 
\begin{equation} 
\Delta j^Z(t,x,i) = j^Z(t,x,i) - \langle  j^Z(t,x,i) \rangle_{\beta,\mu}\quad ,  \qquad Z=N,\, E,\, Q\, ,  
\label{fluc}
\end{equation} 
where $\langle  \cdots \rangle_{\beta,\mu}$ denotes the $\Omega_{\beta, \mu}$-expectation value. 
The noise in the lead $L_i$ can be extracted from the two-point function 
$\langle  \Delta j^Z(t_1,x_1,i)  \Delta j^Z(t_2,x_2,i) \rangle_{\beta,\mu}$, 
which, because of (\ref{fluc}), equals the connected current-current correlator 
\begin{equation} 
\langle  \Delta j^Z(t_1,x_1,i)  \Delta j^Z(t_2,x_2,i) \rangle_{\beta,\mu} = 
\langle  j^Z(t_1,x_1,i)  j^Z(t_2,x_2,i) \rangle_{\beta,\mu}^{\rm conn} \, . 
\label{conn}
\end{equation} 
Since the energy is conserved in the state $\Omega_{\beta, \mu}$, the right hand side of (\ref{conn}) 
depends only on the difference $t_{12}=t_1-t_2$. Accordingly, the noise power at frequency $\nu$ is defined by 
the Fourier transform 
\begin{equation} 
P_i^Z(\beta,\mu,\S;x_1,x_2;\nu) = 
\int_{-\infty}^\infty \rd t \, \e^{\ri \nu t}\, \langle  j^Z(t,x_1,i)  j^Z(0,x_2,i) \rangle_{\beta,\mu}^{\rm conn} \, . 
\label{nunoise}
\end{equation} 
Following \cite{ML-92}-\cite{BB-00}, we focus in what follows on the {\it zero frequency noise power} 
\begin{equation} 
P_i^Z(\beta,\mu,\S) = \lim_{\nu \to 0^+} P_i^Z(\beta,\mu;x_1,x_2;\nu) \, , 
\label{0noise}
\end{equation} 
which has the remarkable feature to be $x_{1,2}$-independent \cite{Mintchev:2011mx}. 

The particle current (\ref{curr1}) satisfies at the junction $x=0$ the Kirchhoff rule  
\begin{equation}
j^N(t,0,1)+j^N(t,0,2) = 0 \, , 
\label{k1}
\end{equation} 
which implies 
\begin{equation} 
P_1^N(\beta,\mu,\S) = P_2^N(\beta,\mu,\S) \equiv P^N (\beta,\mu,\S)\, . 
\label{pn1}
\end{equation} 
Differently from $j^N$, in general the heat current (\ref{hc}) does not satisfy the Kirchhoff 
rule.\footnote{This is the origin of the energy transmutation in the junction described in \cite{MSS-14}.} 
Therefore, $P^Q_1$ and $P^Q_2$ are generally different. Nevertheless, they are 
simply related. In fact, one can obtain $P^Q_2$ from $P^Q_1$ and vice versa by the exchange operation 
$\beta_1 \leftrightarrow \beta_2$ and $\mu_1 \leftrightarrow \mu_2$. For this reason we focus below 
on $P^Q_1$, setting for simplicity 
\begin{equation}
P^Q(\beta,\mu,\S) \equiv P^Q_1(\beta,\mu,\S)\, . 
\label{n1}
\end{equation} 

In what follows we will first derive and study in detail the noises $\pnf$ and $\hnf$ 
for fermions. Afterwards we will 
summarize the results for bosons and discuss the effect of the statistics. 

\bigskip 

\section{Fermionic particle noise} 
\medskip

Using the two-point $j^N$-correlation function (\ref{A3}) in the representation $\H^+_{\rm LB}$ 
(see appendix A), after some algebra one gets 
\begin{equation}
\pnf(\beta,\mu,\tau) = \tau^2 A^N_{\rm f}(\beta,\mu) + \tau (1-\tau)B^N_{\rm f}(\beta,\mu) \, , 
\label{pn2}
\end{equation}
$\tau = |\S_{12}|^2 = |\S_{21}|^2$ being the {\it transmission probability} and 
\begin{equation} 
A^N_{\rm f}(\beta,\mu) =  \int_{0}^{\infty} \frac{dk}{2\pi } \frac{k}{m} \left [d_1(k) + d_2(k)-d_1^2(k) -d_2^2(k) \right ] \, , 
\label{pn3}
\end{equation}
\begin{equation} 
B^N_{\rm f}(\beta,\mu) =  \int_{0}^{\infty} \frac{dk}{2\pi } \frac{k}{m} \left [d_1(k) + d_2(k)-2 d_1(k) d_2(k) \right ] \, ,\quad
\label{pn4}
\end{equation}
were 
\begin{equation} 
d_i(k) = \frac{\e^{-\beta_i \left [\omega (k) -\mu_i\right ]}}
{1+ \e^{-\beta_i \left [\omega (k) -\mu_i \right ]}} \, , \qquad i=1,2\, 
\label{pn5} 
\end{equation} 
is the Fermi distribution in the heat reservoir $R_i$. The behavior of the particle noise as a 
function of the heat bath parameters $(\beta_i,\mu_i)$ and transmission 
probability $\tau$ is fully described by (\ref{pn2}-\ref{pn5}). We stress that $\pnf$ is well defined for 
$\beta_{1,2}> 0$ and on the whole $\mu_{1,2}$-plane. 

\bigskip 

\subsection{General properties}
\medskip

Eq. (\ref{pn2}) provides an useful representation of $\pnf$. Indeed, 
from the explicit form of the integrands of $A^N_{\rm f}$ and $B^N_{\rm f}$ 
one can easily deduce that 
\begin{equation} 
B^N_{\rm f}(\beta,\mu) \geq A^N_{\rm f}(\beta,\mu) \geq 0\, , 
\label{pn6}
\end{equation} 
which, combined with $0\leq \tau\leq1$, implies that the particle noise is non-negative, 
\begin{equation} 
\pnf(\beta,\mu,\tau) \geq 0\, . 
\label{pn7}
\end{equation} 
The lower bound (\ref{pn7}) is actually a direct consequence of the positivity of the scalar product in the 
Hilbert space $\H_{\rm LB}$. 

\begin{figure}[h]
\begin{center}
\begin{picture}(80,100)(80,25) 
\includegraphics[scale=0.75]{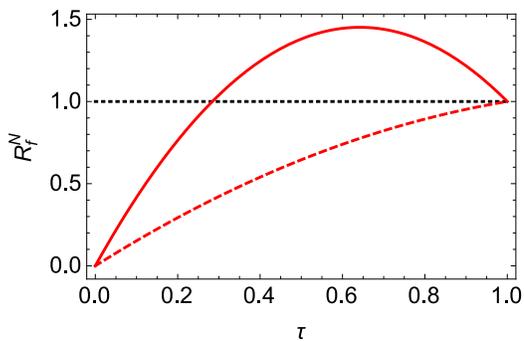}
\end{picture} 
\end{center}
\caption{Plot of the ratio $R^N_{\rm f}=\pnf(\beta,\mu,\tau)/\pnf(\beta,\mu,1)$ for $2A^N_{\rm f} \geq B^N_{\rm f} \geq A^N_{\rm f}$ (dashed line) and $B^N_{\rm f}>2A^N_{\rm f}$ (continuos line).} 
\label{fig2}
\end{figure}

Concerning the $\tau$-dependence, we first observe that (\ref{pn6}) implies that $\pnf$ is a concave function of $\tau$. 
On physical grounds one might be tempted to believe that $\pnf$ is maximal at maximal transmission $\tau_m^N=1$. 
This is indeed the case only in the range of heat bath parameters for which $2 A^N_{\rm f} \geq B^N_{\rm f} \geq A^N_{\rm f}$. 
If instead $B^N_{\rm f}>2A^N_{\rm f}$, the noise $\pnf$ reaches its maximum for some $\tau^N_m <1$. This behavior is 
illustrated in Fig. \ref{fig2}. 

Let us investigate now the dependence of $\pnf$ on the chemical potentials. In order to make contact with 
the previous work \cite{ML-92}-\cite{BB-00}, we set 
\begin{equation} 
\beta_1=\beta_2 \equiv \beta \in \RR_+\, , \qquad \mupm = \frac{1}{2}(\mu_1\pm \mu_2)\in \RR\, , 
\label{pn8} 
\end{equation} 
$\mup$ being the average charge density of the two heat baths. 
The $k$-integration in (\ref{pn3}, \ref{pn4}) can be performed exactly (see appendix B) and one finds 
\begin{eqnarray} 
\pnf(\beta, \mum,\mup,\tau) = 
\frac{\tau}{2\pi \beta} \Biggl\{\frac{\tau \left [1+ 2 \e^{\beta(\mum +\mup)} +  \e^{2\beta \mum }\right ]} 
{\left [1+\e^{\beta(\mum +\mup)}\right ] \left [1+\e^{\beta(\mum -\mup)}\right ]} + \qquad 
\nonumber \\ 
(1-\tau) (\beta \mum )\coth (\beta \mum) + 
(1-\tau) \coth (\beta \mum) \ln \left [\frac{1+ \e^{\beta(\mum +\mup)}}{\e^{\beta \mup } + 
\e^{\beta \mum }}\right ] \Biggr \} \, . 
\label{pn9}
\end{eqnarray} 
For $\mup =0$ this expression significantly simplifies to 
\begin{equation} 
\pnf(\beta, \mum,0,\tau) = 
\frac{\tau}{2\pi \beta} \left [ \tau  + 
(1-\tau) (\beta \mum )\coth (\beta \mum) \right ] \, , 
\label{pn10}
\end{equation} 
which is precisely the result (with our normalization (\ref{curr1}) of the current) 
reported in \cite{ML-92}-\cite{BB-00}. For $\mum=0$ one gets instead 
\begin{equation} 
\pnf(\beta, 0,\mup,\tau) = 
\frac{\tau \e^{\beta \mup}}{\pi \beta \left (1+\e^{\beta \mup}\right )} \, . 
\label{pn10n}
\end{equation} 
Because of (\ref{pn8}) in this case $\mu_1=\mu_2$ and (\ref{pn10n}) describes therefore 
the equilibrium current fluctuations. In the limit $\mup \to 0$ these fluctuations lead to the Johnson-Nyquist law 
\begin{equation} 
\pnf(\beta, 0,0,\tau) = 
\frac{\tau}{2 \pi \beta} \, . 
\label{JN}
\end{equation} 

\begin{figure}[h]
\begin{center}
\begin{picture}(80,100)(80,25) 
\includegraphics[scale=0.75]{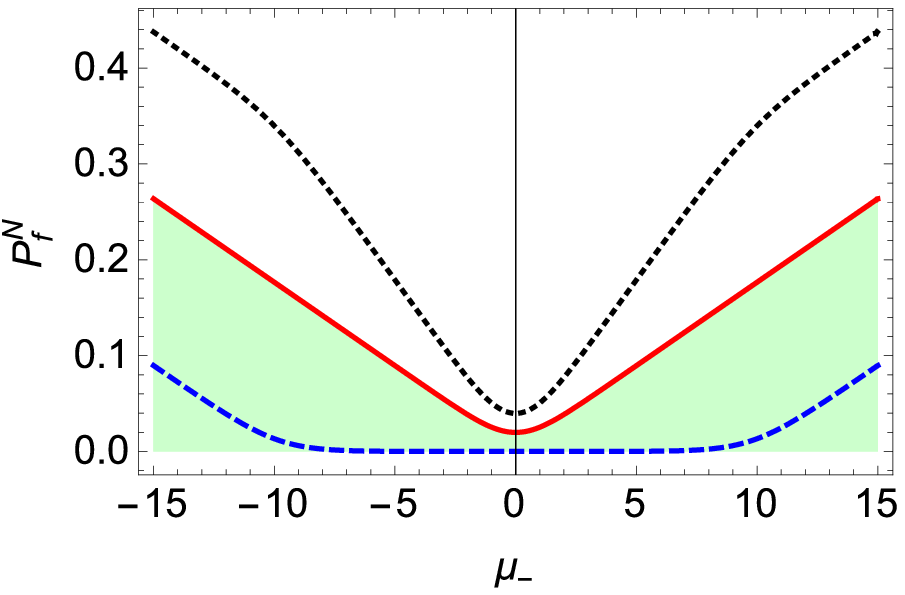}
\end{picture} 
\end{center}
\caption{The noise $\pnf(1,\mum,\mup,1/8)$ (in units of $1/\beta$) as a function 
of $\mum$ for $\mup =10$ (black dotted line), $\mup =0$ (red line) and $\mup =-10$ (blue dashed line).} 
\label{fig3}
\end{figure}

The general expression (\ref{pn9}) has a number of interesting properties. First of all $\pnf$ is 
an even function of $\mum$. To our knowledge, the behavior $\pnf$ as a function of $\mup$ has 
been poorly analyzed previously. It is instructive to fill this gap, because it turns out that the noise 
power depends essentially on the parameter $\mup$. In fact, one can directly verify that 
\begin{equation} 
\partial_{\mup} \pnf(\beta, \mum, \mup, \tau) \geq 0\, , \qquad 
\lim_{\mup \to -\infty} \pnf(\beta, \mum, \mup, \tau) = 0\, , 
\label{pn11} 
\end{equation} 
hold on the whole parameter space ($\beta \in \RR_+,\; \mum \in \RR,\; \tau \in [0,1]$). Therefore, $\pnf$ is 
a {\it positive monotonically increasing} function of the parameter $\mup$, which provides a simple 
mechanism for noise control. In particular,  
for $\mup<0$ ($\mup>0$) the noise $\pnf$ is suppressed (enhanced) with respect to the $\mup=0$ value 
given by (\ref{pn10}) and studied in \cite{ML-92}-\cite{BB-00}. Fig. \ref{fig3} 
displays the particle noise for three different values of $\mup$. The area with green filling represents the 
domain $\mup<0$ of noise reduction. A significant suppression of the noise is observed in the range $\mup < \mum <-\mup$, which is 
in agreement with the fact that $\pnf$ vanishes in the limit $\mup \to -\infty$.  

\begin{figure}[h]
\begin{center}
\begin{picture}(260,100)(80,25) 
\includegraphics[scale=0.75]{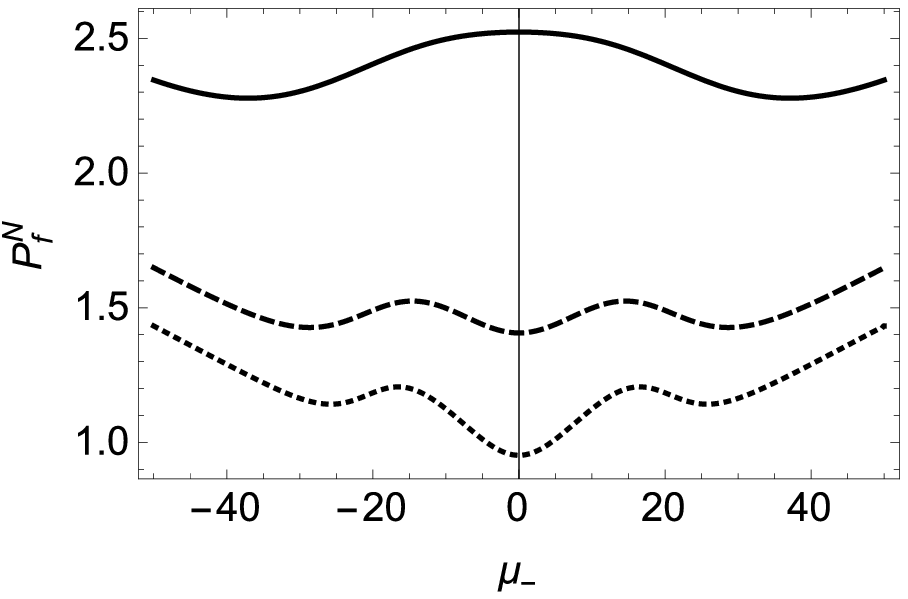} 
\hskip 0.5 truecm
\includegraphics[scale=0.75]{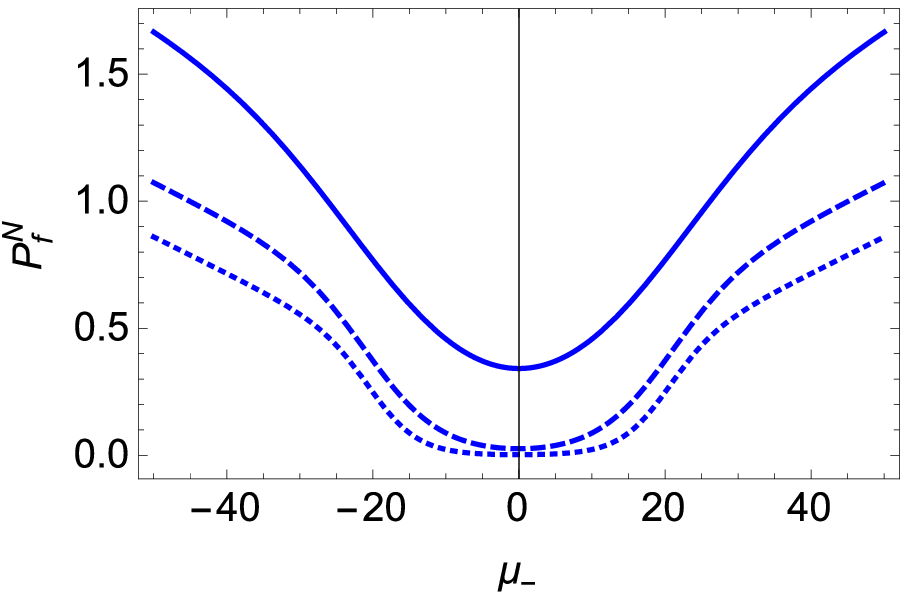}
\end{picture} 
\end{center}
\caption{The left panel represents $\pnf (\beta,\mum,\mup,0.9)$ (in units of $1/\beta$) as a function of $\mum$ for $\mup =20$ 
and $\beta = 0.3$ (black dotted line), $\beta=0.2$ (black dashed line) and $\beta=0.1$ (black continuous line). The right panel 
displays $\pnf(\beta,\mum,\mup,0.9)$ for $\mup =-20$ 
and $\beta = 0.3$ (blue dotted line), $\beta=0.2$ (blue dashed line) and $\beta=0.1$ (blue continuous line).} 
\label{figT}
\end{figure}

The dependence of $\pnf$ on the temperature $1/\beta$ is illustrated in Fig. \ref{figT}. As expected the noise increases 
with increasing the temperature both for $\mup>0$ (left panel) and $\mup<0$ (right panel). 

The shot noise is the zero temperature limit of (\ref{pn9}). Performing this limit, one can assume 
without loss of generality that $\mum \geq 0$ (equivalently $\mu_1\geq \mu_2$). One finds under this condition 
\begin{equation} 
\lim_{\beta \to +\infty} \pnf(\beta, \mum,\mup,\tau) = 
\begin{cases} 
\frac{1}{\pi}\, \tau(1-\tau)\, \mum \, , & \qquad  \mup \geq \mum\, , \\
\frac{1}{2\pi}\, \tau(1-\tau)\, (\mup + \mum )\, , & \qquad  -\mum \leq \mup < \mum\, , \\
0 \, , & \qquad  \mup < -\mum\, . \\ 
\end{cases} 
\label{pn12}
\end{equation} 
which shows the absence of shot noise \cite{L-89} at vanishing 
transmission ($\tau=0$) and at full transmission ($\tau=1$). We also see that the shot noise (\ref{pn12}) is 
a continuos function of $\mup$ with a discontinuous derivative  
in $\mup=\pm \mum$. 

Summarizing, the general result (\ref{pn9}) clearly shows that $\mup$ controls 
the noise power in the LB state $\Omega_{\beta, \mu}$. This feature is of practical relevance because 
the parameter $\mup$ can be directly accessed in experiments. In the next subsection we will relate the 
behavior of $\pnf$ to the non-linearity of the $J^N$-$\mum$ characteristics 
of the junction for $\mup\not=0$. 

\medskip 
\subsection{Particle noise as a function of the current}
\medskip 

In the applications \cite{B-10} to noise spectroscopy it is useful to express the noise (\ref{pn9}) as a function 
of the particle current \cite{L-57,B-86} 
\begin{eqnarray} 
J^N (\beta,\mum,\mup,\tau) &\equiv& \langle j^N(t,x,1)\rangle_{\beta,\mu} = 
\nonumber \\
\tau \int_{0}^{\infty} \frac{dk}{2\pi } \frac{k}{m} \left [d_1(k) - d_2(k)\right ] &=& 
\frac{\tau}{2\pi \beta} 
\ln \left [\frac{1+\e^{\beta(\mup+\mum)}}{1+\e^{\beta(\mup-\mum)}}\right ] \, . 
\label{pn13}
\end{eqnarray} 
As already mentioned in the introduction, the value $\mup=0$ is very special because in this case the current 
\begin{equation}
J^N_0 \equiv J^N (\beta,\mum,0,\tau)= \frac{1}{2\pi} \tau \mum  
\label{pn14}
\end{equation} 
depends linearly on $\mum$. Combining this expression with (\ref{pn10}) one 
gets 
\begin{equation} 
\pnf(\beta, 2\pi J^N_0/\tau,0,\tau) = 
\frac{1}{2\pi \beta} \left [ \tau^2  + 
(1-\tau) 2 \pi \beta J^N_0 \coth \left (2 \pi \beta J^N_0/\tau \right ) \right ] \, , 
\label{snc}
\end{equation} 
which has been tested experimentally in \cite{K-96}. The zero temperature limit of (\ref{snc}) gives in particular 
\begin{equation} 
\lim_{\beta \to +\infty} \pnf(\beta, 2\pi J^N_0/\tau,0,\tau) = (1-\tau)|J^N_0| \, . 
\label{lsnc}
\end{equation} 

\begin{figure}[h]
\begin{center}
\begin{picture}(80,100)(80,25) 
\includegraphics[scale=0.75]{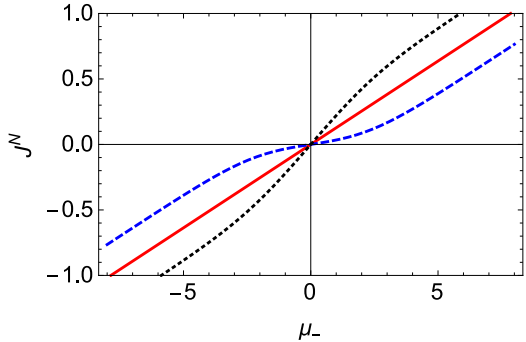}
\end{picture} 
\end{center}
\caption{The current $J^N(1,\mum,\mup,0.8)$ (in units of $1/\beta$) as a function of $\mum$ for $\mup =2$ (black dotted line), 
$\mup =0$ (red line) and $\mup =-2$ (blue dashed line).} 
\label{fig4}
\end{figure}

Our goal here is to generalize (\ref{snc}) to the case $\mup \not=0$. 
It is evident from (\ref{pn13}) that for $\mup \not=0$ the current $J^N$ depends non-linearly on $\mum$,  
which is illustrated by the blue (dashed) and black (dotted) curves in Fig. \ref{fig4}. 
It is worth mentioning that such non-linear dependence has been 
experimentally observed in the double-barrier tunnel junctions studied in \cite{B-95}. In fact, 
the $J^N$-$\mum$ curve \cite{B-95} relative to these devises resembles very much the blue dashed 
line in Fig. \ref{fig4}. Analogous non-linear behavior has been detected in superconducting 
tunnel junctions in \cite{BDD-05}. 

\begin{figure}[h]
\begin{center}
\begin{picture}(260,100)(80,25) 
\includegraphics[scale=0.75]{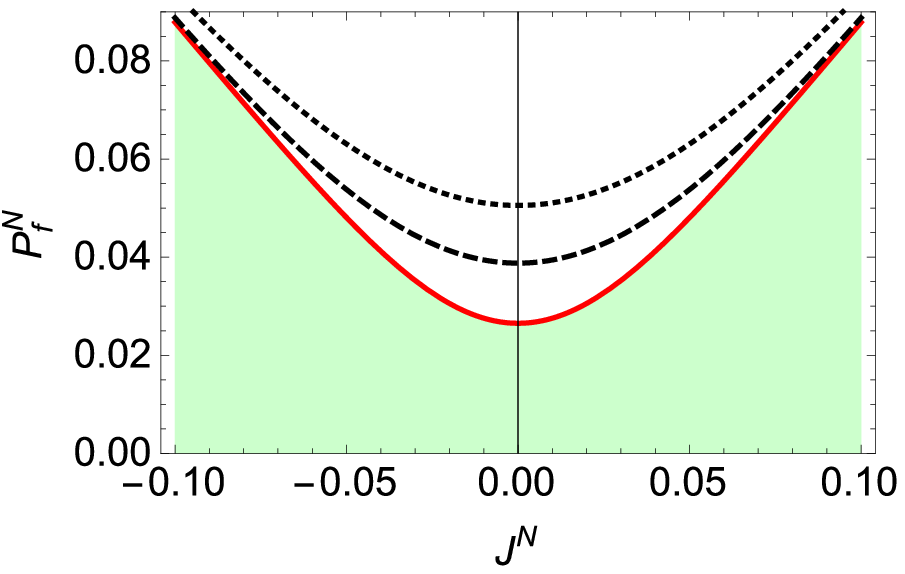} 
\hskip 0.5 truecm
\includegraphics[scale=0.75]{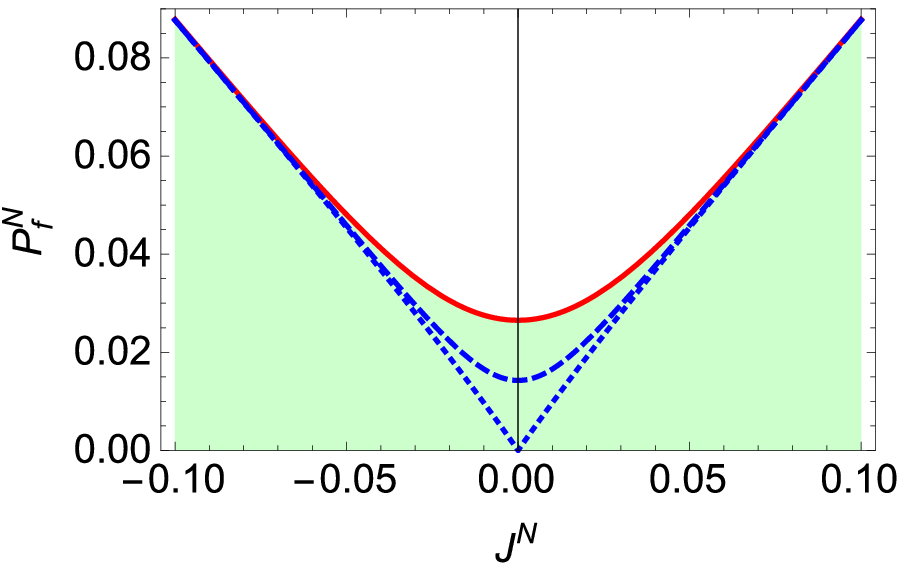}
\end{picture} 
\end{center}
\caption{$\pnf$ (in units of $1/\beta$) as a function of $J^N$ for $\beta=1$ and $\tau =1/6$, represented 
for $\mup =0$ by a red line in both panels. The same function for 
$\mup =1$ (black dashed line) and 
$\mup=3$ (black dotted line) in the left panel, and 
$\mup =-1$ (blue bashed line) and $\mup =-10$ (blue dotted line) in the right panel.} 
\label{fig5}
\end{figure}

Solving (\ref{pn13}) for $\mum$ one gets  
\begin{equation} 
\mum(J^N) = \frac{1}{\beta} \ln\left [\frac{1}{2}\left (\e^{2\pi\beta J^N/\tau} -1\right ) + 
\sqrt {\frac{1}{4}\left (\e^{2\pi\beta J^N/\tau} -1\right )^2 + \e^{2\beta \mup+2\pi\beta J^N/\tau}}\, \right ] - \mup \, . 
\label{pn15}
\end{equation} 
By means of (\ref{pn15}) one can eliminate $\mum$ in favor of $J^N$ in the general formula (\ref{pn9}) and 
obtain the noise power $\pnf (\beta, \mum(J^N),\mup,\tau)$. The dependence of this function 
on $J^N$ is displayed in Fig. \ref{fig5}. The red curve corresponds to $\mup=0$ and describes 
the behavior of (\ref{snc}) studied experimentally for different temperatures and conductances in \cite{K-96}. 
The new black (left panel) and blue (right panel) curves are obtained for $\mup>0$ and 
$\mup<0$ respectively. We see that for the same value of the current $J^N$ flowing in the junction, 
negative (positive) values of $\mup$ reduce (enhance) the noise with respect to $\mup=0$.

Let us observe that according to (\ref{pn15}) the value $J^N=0$ implies $\mum=0$ and 
leads therefore to the equilibrium current fluctuations given by (\ref{pn10n}). 

\bigskip

\subsection{Relaxing the condition of scale invariance}
\medskip 

We conclude our study of the particle noise by considering momentum dependent scattering matrices. 
Scale invariance in the junction is broken in this case and in general  
the $k$-integration cannot be performed in exact and explicit form. Nevertheless, one can establish 
some useful estimates. Let us introduce for this purpose the maximal and 
minimal transmission probabilities 
\begin{equation}
\tau_{\rm min} = \min \{ |\S_{12}(k)|^2\, :\, k\geq 0 \}\, , \qquad 
\tau_{\rm max} = \max \{ |\S_{12}(k)|^2\, :\, k\geq 0 \}\, , 
\label{est1}
\end{equation}
which exist because $0\leq |\S_{12}(k)|^2 \leq 1$ by unitarity. Then $\pnf$ satisfies the following 
inequalities  
\begin{eqnarray}
\pnf(\beta,\mu,\S) &\geq& 
\tau_{\rm min}^2\, A^N_{\rm f}(\beta,\mu) + 
\tau_{\rm min} (1-\tau_{\rm max})\, B^N_{\rm f}(\beta,\mu) \, , 
\label{est2a}\\
\pnf(\beta,\mu,\tau) &\leq& \tau_{\rm max}^2\, A^N_{\rm f}(\beta,\mu) + 
\tau_{\rm max} (1-\tau_{\rm min})\, B^N_{\rm f}(\beta,\mu) \, , 
\label{est2b}
\end{eqnarray}
where $A^N_{\rm f}$ and $B^N_{\rm f}$ are the $\S$-independent integrals (\ref{pn3},\ref{pn4}). 
The estimates (\ref{est2a},\ref{est2b}) follow only from unitarity and in this sense are universal. 

Let us consider at this point the scattering matrices (\ref{S1}) generated by the boundary 
condition (\ref{bc1}). Inserting in (\ref{d3}) the most general $2\times 2$ unitary matrix $\U$, 
one obtains \cite{Mintchev:2011mx} the transmission probability 
\begin{equation} 
|\S_{12}(k)|^2 = |\S_{21}(k)|^2 = \frac{k^2 (\eta_1-\eta_2)^2 \sin^2 (\theta)}{(k^2+\eta_1^2)(k^2+\eta_2^2)} \, , 
\qquad \theta \in [0,2\pi)\, , 
\label{est3}
\end{equation} 
where $\eta_i$ are given by (\ref{d4}) and the angle $\theta$ is the only 
parameter which remains from the four ones characterizing a generic $\U\in U(2)$. From (\ref{est3}) one gets 
\begin{equation} 
\tau_{\rm min} = 0\, , \qquad \tau_{\rm max}(\eta_i,\theta) = 
\frac{(\eta_1-\eta_2)^2 \sin^2 (\theta)}{(|\eta_1|+|\eta_2|)^2}\, .  
\label{est4}
\end{equation} 
The combination of (\ref{est2a},\ref{est2b}) with (\ref{est4}) implies 
\begin{equation} 
0\leq \pnf(\beta,\mu,\S) \leq \tau_{\rm max}^2(\eta_i,\theta)\, A^N_{\rm f}(\beta,\mu) + 
\tau_{\rm max}(\eta_i,\theta) \, B^N_{\rm f}(\beta,\mu) \, , 
\label{est5}
\end{equation} 
which confirms (\ref{pn7}) away from criticality and provides an upper bound 
on the particle noise for the whole family (\ref{S1}) of $k$-dependent scattering matrices.

\bigskip

\section{Fermionic heat noise} 
\medskip 

The above analysis can be extended to the heat current (\ref{hc}). One finds 
\begin{equation}
P_{\rm f}^Q(\beta,\mu,\tau) = \tau^2 A^Q_{\rm f} (\beta,\mu) + \tau (1-\tau) B^Q_{\rm f} (\beta,\mu)\, , 
\label{hn1}
\end{equation}
where 
\begin{equation} 
A^Q_{\rm f}(\beta,\mu) =  \int_{0}^{\infty} \frac{dk}{2\pi } \frac{k}{m}\, 
\left [\omega(k)-\mu_1\right ]^2 \left [d_1(k) + d_2(k)-d_1^2(k) -d_2^2(k) \right ] \, , 
\label{hn2}
\end{equation}
\begin{equation} 
B^Q_{\rm f}(\beta,\mu) =  \int_{0}^{\infty} \frac{dk}{2\pi } \frac{k}{m}\, 
\left [\omega(k)-\mu_1\right ]^2 \left [d_1(k) + d_2(k)-2 d_1(k) d_2(k) \right ] \, .\quad
\label{hn3}
\end{equation} 
Analogously to the particle noise, the integrals (\ref{hn2},\ref{hn3}) satisfy 
\begin{equation} 
B^Q_{\rm f}(\beta,\mu) \geq A^Q_{\rm f}(\beta,\mu) \geq 0\, , 
\label{hn4}
\end{equation} 
implying that the heat noise is non-negative, 
\begin{equation} 
P^Q_{\rm f}(\beta,\mu,\tau) \geq 0\, . 
\label{hn5}
\end{equation} 

The $\tau$-dependence of $\hnf$ is similar to that of the particle noise. In fact, $\hnf$ is maximal at $\tau^Q_m=1$ if 
$2A^Q_{\rm f}\geq B^Q_{\rm f} \geq A^Q_{\rm f}$. If instead $B^Q_{\rm f}>2A^Q_{\rm f}$, 
the heat noise $\hnf$ reaches its maximum for $\tau^Q_m<1$. 
The dependence of the heat noise (\ref{hn1}) on $\beta$ and $\mupm$ is much more 
involved. For the sake of conciseness 
we report the explicit form of $\hnf$ in appendix C, summarizing here its basic properties. 
One can deduce from (\ref{C5}) that 
\begin{equation} 
\partial_{\mup} \hnf(\beta, \mum, \mup, \tau) \geq 0\, , \qquad 
\lim_{\mup \to -\infty} \hnf(\beta, \mum, \mup, \tau) = 0\, , 
\label{hn6} 
\end{equation} 
which implies that $\hnf$ is an increasing function of $\mup$. Differently from $\pnf$ however, $\hnf$ is not 
symmetric in $\mum$.  

\begin{figure}[h]
\begin{center}
\begin{picture}(80,100)(80,25) 
\includegraphics[scale=0.75]{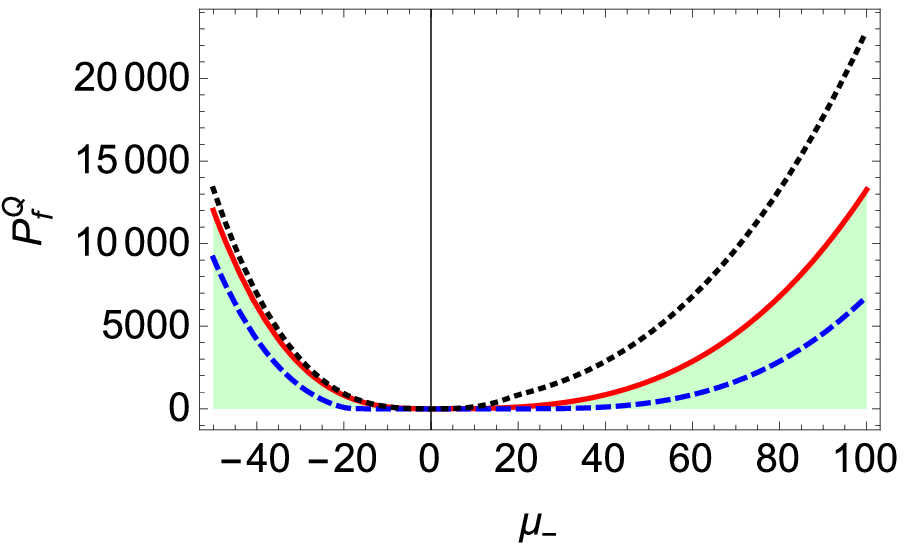}
\end{picture} 
\end{center}
\caption{The noise $\hnf(1,\mum,\mup,1/2)$ (in units of $1/\beta$) as a function of $\mum$ for $\mup =20$ 
(black dotted line), $\mup =0$ (red line) and $\mup =-20$ (blue dashed line).} 
\label{fig6}
\end{figure} 

The plots in Fig. \ref{fig6} give an idea about the behavior of $\hnf$ for different values of $\mup$. 
The heat noise power is reduced for negative $\mup$ and enhanced for positive $\mup$, exactly 
like the particle noise $\pnf$. 

The zero temperature limit of (\ref{C5}) gives (for $\mum \geq 0$): 
\begin{equation} 
\lim_{\beta \to \infty} \hnf(\beta, \mum,\mup,\tau) = 
\begin{cases} 
\frac{4}{3\pi}\, \tau(1-\tau)\, \mum^3 \, , & \qquad  \mup \geq \mum\, , \\
\frac{1}{6\pi}\, \tau(1-\tau)\, (\mup + \mum )^3\, , & \qquad  -\mum \leq \mup < \mum\, , \\
0 \, , & \qquad  \mup < -\mum\, . \\ 
\end{cases} 
\label{hn7}
\end{equation} 
At equilibrium ($\mum=0$) one finds  
\begin{equation} 
\hnf(\beta, 0,\mup,\tau) = 
\frac{\tau \left \{\beta^2 \mup^2 \e^{\beta \mup} -2 \left (1+\e^{\beta \mup}\right )
\left [\beta \mup \ln \left (1+\e^{\beta \mup}\right ) 
+\li_2 \left (-\e^{\beta \mup}\right )\right ] \right \}}
{\pi \beta^3 \left (1+\e^{\beta \mup}\right )}\, . 
\label{hn8}
\end{equation} 
Setting in addition $\mup=0$ one has 
\begin{equation} 
\hnf(\beta, 0,0,\tau) = 
\frac{\pi \tau } {6 \beta^3}\, , 
\label{hn9}
\end{equation} 
which is the counterpart of the Johnson-Nyquist law concerning the heat noise.

\bigskip 

\section{Bosonic particle fluctuations} 
\medskip 

\begin{figure}[h]
\begin{center}
\begin{picture}(80,100)(80,25) 
\includegraphics[scale=0.75]{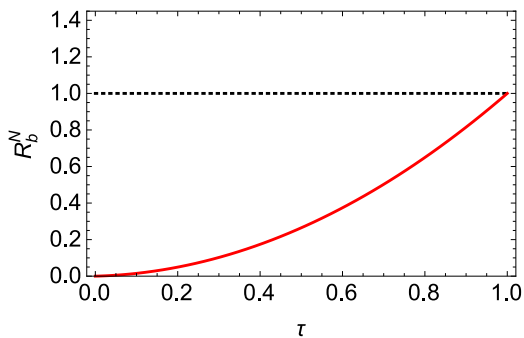}
\end{picture} 
\end{center}
\caption{Plot of the ratio $R^N_{\rm b}=\pnb(\beta,\mu,\tau)/\pnb(\beta,\mu,1)$.} 
\label{fig7}
\end{figure}

In the bosonic case one gets from (\ref{conn})-(\ref{0noise}) the particle noise 
\begin{equation}
\pnb(\beta,\mu,\tau) = \tau^2 A^N_{\rm b}(\beta,\mu) + \tau (1-\tau)B^N_{\rm b}(\beta,\mu) \, , 
\label{pnb2}
\end{equation}
where now 
\begin{equation} 
A^N_{\rm b}(\beta,\mu) =  \int_{0}^{\infty} \frac{dk}{2\pi } \frac{k}{m} \left [d_1(k) + d_2(k)+d_1^2(k) +d_2^2(k) \right ] \, , 
\label{pnb3}
\end{equation}
\begin{equation} 
B^N_{\rm b}(\beta,\mu) =  \int_{0}^{\infty} \frac{dk}{2\pi } \frac{k}{m} \left [d_1(k) + d_2(k)+2 d_1(k) d_2(k) \right ] \, ,\quad
\label{pnb4}
\end{equation} 
and $d_i(k)$ is the Bose distribution 
\begin{equation} 
d_i(k) = \frac{\e^{-\beta_i \left [\omega (k) -\mu_i\right ]}}
{1- \e^{-\beta_i \left [\omega (k) -\mu_i \right ]}} \, , \qquad i=1,2\, . 
\label{pnb5} 
\end{equation} 
A new characteristic feature of the bosonic case is the presence of singularities in the integrands of (\ref{pnb3},\ref{pnb4}) at 
$k=\sqrt{2 m \mu_i}$. In order to avoid them and deal with meaningful expressions, we impose 
\begin{equation}
\mu_1<0\quad {\rm and}\quad \mu_2<0 \quad  \Longleftrightarrow \quad \mup <0\quad {\rm and}\quad \mup < \mum <-\mup \, . 
\label{pnb6}
\end{equation}
This assumption shifts the singularities away from the range of integration. We stress that all our 
results about bosonic systems hold only 
in the open cone $\C=\{\mup <0,\, \mup < \mum  <-\mup\}$ defined by (\ref{pnb6}). 
The integrals (\ref{pnb3},\ref{pnb4}) are well defined there and satisfy 
\begin{equation} 
A^N_{\rm b}(\beta,\mu) \geq B^N_{\rm b}(\beta,\mu) \geq 0\, . 
\label{pnb7}
\end{equation} 
The bosonic noise power $\pnb$ is therefore non-negative like the fermionic one. However, differently from the fermionic 
case, $\pnb$ is a convex function (see Fig. \ref{fig7}) of the transmission probability $\tau$. This fact follows directly from 
(\ref{pnb2},\ref{pnb7}) and provides a possible tool for detecting the statistics in noise spectroscopy. 

The cone $\C$ represents the common domain where both 
$\pnf$ and $\pnb$ are well defined and can be therefore compared. From the inequalities (\ref{B6},\ref{B7}) in 
appendix B it follows that for the same heat bath parameters the bosonic noise always exceeds the fermionic one, 
\begin{equation}
\pnb(\beta, \mu)  \geq \pnf (\beta, \mu)\, ,\qquad (\mup, \mum)\in \C\, . 
\label{n2}
\end{equation}
This remarkable feature illustrates the deep role \cite{L-98} of the exclusion principle in reducing the  
quantum noise in the domain $\C$, where the bosonic transport is well defined. 
The result (\ref{n2}) can find interesting applications in quantum transport. 

For $\beta_1=\beta_2\equiv \beta$ the explicit expression of $\pnb$ in $\C$ is (see appendix B) 
\begin{eqnarray} 
\pnb(\beta, \mum,\mup,\tau) = 
\frac{\tau}{2\pi \beta} \Biggl\{\frac{\tau \left [2 \e^{\beta(\mum +\mup)} - \e^{2\beta \mum }-1\right ]} 
{\left [1-\e^{\beta(\mum +\mup)}\right ] \left [1-\e^{\beta(\mum -\mup)}\right ]} - \qquad 
\nonumber \\ 
(1-\tau) (\beta \mum )\coth (\beta \mum) -
(1-\tau) \coth (\beta \mum) \ln \left [\frac{1- \e^{\beta(\mum +\mup)}}{\e^{\beta \mum } - 
\e^{\beta \mup }}\right ] \Biggr \} \, . 
\label{pnb8}
\end{eqnarray} 
The presence of singularities outside of the open cone $\C$ is manifest in (\ref{pnb8}), which 
diverges on the boundary $\mum=\pm \mup$ and on the tip $\mup=0$ of $\C$. 
The lines $\mum=0$ with $\mup\not=0$ belong to $\C$ and 
\begin{equation} 
\pnb(\beta, 0,\mup,\tau) = 
\frac{\tau \e^{\beta \mup}}{\pi \beta \left (1-\e^{\beta \mup}\right )} \, , 
\label{pn9b}
\end{equation} 
which describes the equilibrium bosonic particle fluctuations. 

\begin{figure}[h]
\begin{center}
\begin{picture}(80,100)(80,25) 
\includegraphics[scale=0.75]{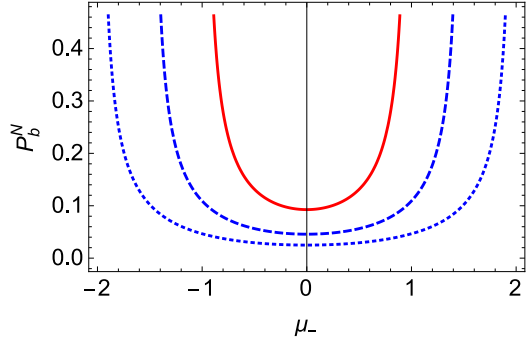}
\end{picture} 
\end{center}
\caption{The noise $\pnb(1,\mum,\mup,1/2)$ (in units of $1/\beta$) 
as a function of $\mum$ for $\mup =-1$ (red line), $\mup =-1.5$ (dashed blue line) and $\mup =-2$ (dotted blue line).} 
\label{fig8}
\end{figure}

Like for fermions, $\pnb$ is a symmetric function of $\mum$, which increases monotonically 
with $\mup \in (-\infty,\, 0)$ and vanishes in the limit $\mup \to -\infty$. Fig. \ref{fig8} illustrates 
the behavior of $\pnb$ in the variable $\mum$ for three different values of $\mup$. 
With the substitution $A^N_{\rm f}\mapsto A^N_{\rm b}$ and $B^N_{\rm f}\mapsto B^N_{\rm b}$ the bounds 
(\ref{est2a},\ref{est2b},\ref{est5}) hold for non-critical bosonic junctions as well.

\bigskip 
\section{Multi-terminal junctions} 
\medskip 

We generalize here the above results to junctions with $n>2$ leads, concentrating on the particle 
noise power for fermions. Plugging (\ref{A3}) in (\ref{nunoise}) one finds in the $i$-th terminal  
\begin{eqnarray}
\pnf(\beta,\mu,\S;i) = 
\qquad \qquad \qquad \qquad \qquad \qquad 
\nonumber \\
\int_0^\infty \frac{\rd k}{2\pi} \frac{k}{m} \Biggl \{(1-2|\S_{ii}|^2) 
d_i(k)\left [1-d_i(k)\right ] + 
\sum_{l=1}^n |\S_{il}|^2 d_l(k)-\left [\sum_{l=1}^n |\S_{il}|^2 d_l(k) \right ]^2\Biggr \}\, . 
\label{mt1}
\end{eqnarray}
In order to illustrate the physics behind this complicated formula, which depends on many parameters, 
it is instructive to consider the case $n=3$. One can interpret the lead $i=3$ as a probe terminal and study for instance 
its influence on the noise in the lead $i=1$. Setting  
$\tau = |\S_{12}|^2$ and $\sigma = |\S_{13}|^2$ one gets from (\ref{mt1}) 
\begin{equation} 
\pnf(\beta,\mu,\S;1) = \pnf(\beta,\mu;\tau)_{(1,2)} + \pnf(\beta,\mu;\sigma)_{(1,3)} +  \pnf(\beta,\mu;\tau,\sigma)_{(1,2,3)}\, ,
\label{mt2}
\end{equation}
where $\pnf(\beta,\mu;\tau)_{(1,2)}$ coincides with the two-lead expression (\ref{pn2}), $\pnf(\beta,\mu;\sigma)_{(1,3)}$ is 
still (\ref{pn2}) but with $\beta_2 \longmapsto \beta_3$, $\mu_2 \longmapsto \mu_3$, 
$\tau \longmapsto \sigma$ and 
$\pnf(\beta,\mu;\tau,\sigma)_{(1,2,3)}$ is a term mixing all three reservoirs with the form 
\begin{equation} 
\pnf(\beta,\mu;\tau,\sigma)_{(1,2,3)} = -2 \tau \sigma \int_0^\infty \frac{\rd k}{2\pi} \frac{k}{m} 
\left [d_1(k)-d_2(k)\right ] \left [d_1(k)-d_3(k)\right ]\, . 
\label{mt3}
\end{equation} 
As expected, (\ref{mt2}) is symmetric under the exchange $2 \leftrightarrow 3$.  
For $\beta_1=\beta_2=\beta_3 \equiv \beta$ the $k$-integrals can be computed explicitly. One has
\begin{eqnarray} 
\pnf(\beta,\mu,\S;1) = \pnf(\beta,\mum,\mup, \tau) + 
\qquad \qquad \qquad \qquad 
\nonumber \\
\pnf(\beta,\mup+\mum-\mu_3,\mup+\mum+\mu_3;\sigma) +  
\pnf(\beta, \mup +\mum,\mup-\mum,\mu_3;\tau,\sigma)\, ,
\label{mt4}
\end{eqnarray}
where the first two terms are given by (\ref{pn9}) and 
\begin{eqnarray} 
\pnf(\beta,\mu_1,\mu_2,\mu_3;\tau,\sigma)= 
\qquad \qquad \qquad \qquad \qquad \qquad 
\nonumber \\
\frac{\tau \sigma}{\pi \beta} 
\Biggl [\frac{\e^{\beta \mu_1}-\left (1+\e^{\beta \mu_1}\right )\ln \left (1+\e^{\beta \mu_1} \right )}
{1+\e^{\beta \mu_1}}  
+ \frac{\e^{\beta \mu_1} \ln \left (1+\e^{\beta \mu_2} \right ) - 
\e^{\beta \mu_2} \ln \left (1+\e^{\beta \mu_1} \right )}{\e^{\beta \mu_1} - \e^{\beta \mu_2}} \qquad 
\nonumber \\ 
\frac{\e^{\beta \mu_1} \ln \left (1+\e^{\beta \mu_3} \right ) - 
\e^{\beta \mu_3} \ln \left (1+\e^{\beta \mu_1} \right )}{\e^{\beta \mu_1} - \e^{\beta \mu_3}}
-\frac{\e^{\beta \mu_2} \ln \left (1+\e^{\beta \mu_3} \right ) - 
\e^{\beta \mu_3} \ln \left (1+\e^{\beta \mu_2} \right )}{\e^{\beta \mu_2} - \e^{\beta \mu_3}}\Biggr ]\, .
\label{mt5}
\end{eqnarray}

\begin{figure}[h]
\begin{center}
\begin{picture}(80,100)(80,25) 
\includegraphics[scale=0.75]{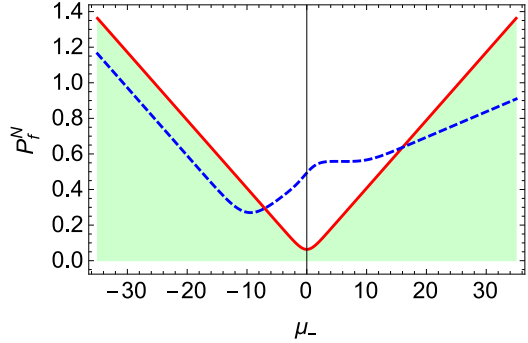}
\end{picture} 
\end{center}
\caption{The noise $\pnf(1,\mum,\mup=0,\mu_3=10,\tau=0.4,\sigma)$ (in units of $1/\beta$) 
as a function of $\mum$ for $\sigma =0$ (red line) and $\sigma =0.5$ (dashed blue line).} 
\label{fig9}
\end{figure}

If one isolates the probe terminal by setting $\sigma =0$, eq.(\ref{mt4}) exactly reproduces the noise (\ref{pn9}) of the 
two-terminal junction. This feature is illustrated by the continuous (red) curve in Fig. \ref{fig9}. For $\sigma \not=0$ 
instead, the probe terminal affects the noise as shown by the dashed (blue) line in the same figure. We see that 
turning on a probe terminal reduces the noise outside of some neighborhood of $\mum=0$. The 
addition of probe terminals represents therefore a viable mechanism for decreasing the noise.

\bigskip 
\section{Outlook and conclusions} 
\medskip 

We studied in this paper some non-linear effects in the particle and heat current fluctuations 
in scale invariant quantum junctions in the Landauer-B\"uttiker non-equilibrium 
steady state. These effects are associated to the non-linear dependence of the particle current $J^N$ 
on the chemical potential difference $\mum$. In the case 
with two heat baths our findings can be summarized as follows: 

(i) Depending on the value of the sum $\mup$ of the chemical potentials, the non-linear effects 
lead to reduction ($\mup<0$) or enhancement ($\mup>0$) of both particle and heat noise for fermions; 

(ii) In the bosonic case the quantum transport is well defined for $\mup <0$ and decreases with 
decreasing of $\mup$. In the common domain $\mup<0$ and with the same values for the heat bath 
parameters the bosonic particle noise always exceeds the fermionic one; 

(iii) Another difference between fermions and bosons concerns the behavior of the particle noise as a function of the 
transmission probability in the junction. For fermions this function is concave, being instead convex for bosons.  

In this context we established an universal upper bound on the particle fluctuations away of criticality. 
We considered also multi-terminal junctions, investigating how the chemical potential of a probe terminal affects 
the particle noise. 

The above results have been extracted from the two-point current correlation functions. 
The theoretical progress in the study of mesoscopic systems has led to the concept \cite{LLL-96}-\cite{LS-14} 
of full counting statistics, in which a more detailed information about the non-equilibrium particle transfer is obtained 
from the $n$-point current correlators with $n>2$.  We expect that the extension of our work 
to this case will shed additional light on the non-linear character of the transport in quantum junctions.

\bigskip



\appendix
\bigskip 
\section{The Landauer-B\"uttiker representations $\H^\pm_{\rm LB}$} 
\medskip 

{}For a detailed construction of the LB representation $\H^\pm_{\rm LB}$ of the algebra $\A_\pm$ we refer to 
\cite{Mintchev:2011mx}. The two-point function $\langle a_j^*(p)a_i(k)\rangle_{\beta, \mu}$ in $\H^\pm_{\rm LB}$ 
is given by  
\begin{eqnarray} 
\langle a_j^*(p)a_i(k)\rangle_{\beta, \mu} = 
2\pi \delta (k-p)\left [\theta(k)\delta_{ij}d^\pm_i(k) + 
\theta(-k)\sum_{l=1}^2 \S_{il}\, d^\pm_l(-k)\, \S^*_{lj}\right ]  \qquad 
\nonumber \\
+2\pi \delta (k+p)\left [\theta(k) d^\pm_i(k) \S^*_{ij} + \theta(-k)\S_{ij} d^\pm_j(-k) \right ] \, , \quad \; \; 
\label{A1}
\end{eqnarray} 
were $d^+_i(k)$ is the Fermi distribution (\ref{pn5}) and $d^-_i(k)$ is the Bose distribution (\ref{pnb5}). 
The explicit form of $\langle a_i(k)a_j^*(p)\rangle_{\beta, \mu}$ is 
obtained from (\ref{A1}) by the substitution 
\begin{equation}
d^\pm_i(k) \longmapsto 1\mp d^\pm_i(k) \, . 
\label{A2}
\end{equation}
As well known \cite{BR}, employing the relation (\ref{rta2}) one can express  
a generic $n$-point correlation function as a polynomial of the two-point 
correlators $\langle a_j^*(p)a_i(k)\rangle_{\beta, \mu}$ and $\langle a_i(k)a_j^*(p)\rangle_{\beta, \mu}$. 
Using this property and (\ref{A1}) one finds   
\begin{eqnarray}
\langle j^N(t_1,x_1,i_1) j^N(t_2,x_2,i_2) \rangle_{\beta, \mu}^{\rm conn}  = \qquad \qquad \qquad \qquad \qquad 
\nonumber \\
\frac{-1}{4m^2} \int^{\infty}_0 \frac{\rd k_1}{2\pi}  \int^{\infty}_0 \frac{\rd k_2}{2\pi} \e^{\ri t_{12} [\omega(k_1) - \omega(k_2)]} 
\sum_{l,m=1}^n d^\pm_l (k_1) [1\mp d^\pm_m (k_2)] \qquad \qquad \quad 
\nonumber \\
\times \Bigl \{ \chi^*_{li_1}(k_1;x_1) \left [\der_{x} \chi_{i_1 m}\right ](k_2;x_1) - 
\left [\der_{x} \chi^*_{l i_1}\right ](k_1;x_1) \chi_{i_1m}(k_2;x_1) \Bigr \} \qquad \qquad \; \; 
\nonumber \\
\times \Bigl \{ \chi^*_{mi_2}(k_2;x_2) \left [\der_{x} \chi_{i_2 l}\right ](k_1;x_2) - 
\left [\der_{x} \chi^*_{m i_2}\right ](k_2;x_2) \chi_{i_2l}(k_1;x_2) \Bigr \}\, ,  \qquad \qquad 
\label{A3}
\end{eqnarray}
where $\chi$ is the matrix
\begin{equation}
\chi(k;x) = \e^{-\ri k x}\, \II + \e^{\ri k x}\, \S\, . 
\label{A4}
\end{equation}
The correlation function $\langle j^E(t_1,x_1,i_1) j^E(t_2,x_2,i_2) \rangle_{\beta, \mu}^{\rm conn}$ 
is obtained from (\ref{A3}) by inserting the factor $[\omega(k_1) + \omega(k_2)]^2$ in the integrand.

\bigskip 
\section{Particle noise integrals} 
\medskip 

Adopting the variables 
\begin{equation}
y=\e^{-\beta_2 k^2/2m}\, ,\quad r=\beta_1/\beta_2\, ,\quad a_i = \e^{-\beta_i\mu_i}\, , \quad i=1,2\, , 
\label{B1}
\end{equation}
the integrals (\ref{pn3},\ref{pn4}) and (\ref{pnb3}, \ref{pnb4}) can be written in the form:
\begin{eqnarray} 
A^N_{\rm f} &=& \frac{1}{2\pi \beta_2} \int_0^1 \rd y \left [\frac{a_1y^{r-1}}{(y^r+a_1)^2} +  \frac{a_2}{(y+a_2)^2}\right ]\, ,
\label{B2}\\
B^N_{\rm f} &=& \frac{1}{2\pi \beta_2} \int_0^1 \rd y\, \frac{a_1+a_2y^{r-1}}{(y^r+a_1)(y+a_2)}\, , 
\label{B3}\\
A^N_{\rm b} &=& \frac{1}{2\pi \beta_2} \int_0^1 \rd y \left [\frac{a_1y^{r-1}}{(y^r-a_1)^2} +  \frac{a_2}{(y-a_2)^2}\right ]\, ,
\label{B4} \\
B^N_{\rm b} &=& \frac{1}{2\pi \beta_2} \int_0^1 \rd y\, \frac{a_1+a_2y^{r-1}}{(y^r-a_1)(y-a_2)}\, . 
\label{B5}
\end{eqnarray} 
These integrals are well defined for $r>0$ and $\mu_i<0$ ($a_i>1$). Moreover, in this domain 
\begin{eqnarray} 
A^N_{\rm b}-A^N_{\rm f} &=& \frac{1}{\pi \beta_2} \int_0^1 \rd y \left [\frac{2a_1^2y^{2r-1}}{(y^{2r}-a_1^2)^2} +  
\frac{2a_2^2y}{(y^2-a_2^2)^2}\right ]\geq 0\, ,
\label{B6}\\
B^N_{\rm b}-B^N_{\rm f} &=& \frac{1}{\pi \beta_2} \int_0^1 \rd y\, \frac{y(a_1+a_2y^{r-1})^2}{(y^{2r}-a_1^2)(y^2-a_2^2)}\geq 0\, ,
\label{B7}
\end{eqnarray}
since the integrands in (\ref{B6},\ref{B7}) are non-negative. 

Performing the integration in (\ref{B2},\ref{B4}) one finds 
\begin{eqnarray} 
A^N_{\rm f} &=& \frac{1}{2\pi \beta_2} \left [\frac{1}{r(a_1+1)} +  \frac{1}{(a_2+1)}\right ]\, ,
\label{B8}\\
A^N_{\rm b} &=& \frac{1}{2\pi \beta_2} \left [\frac{1}{r(a_1-1)} +  \frac{1}{(a_2-1)}\right ]\, ,
\label{B9} 
\end{eqnarray} 
The explicit form of  (\ref{B3},\ref{B5}) for generic $r$ is not known. For $r=1$ ($\beta_1=\beta_2\equiv \beta$) one has
\begin{eqnarray} 
B^N_{\rm f} &=& \frac{1}{2\pi \beta}\, \frac{(a_1+a_2)}{(a_1-a_2)}\ln\left [\frac{a_1a_2 + a_1}{a_1a_2+a_2}\right]\, , 
\label{B10}\\
B^N_{\rm b} &=& \frac{1}{2\pi \beta}\, \frac{(a_1+a_2)}{(a_1-a_2)}\ln\left [\frac{a_1a_2 -a_2}{a_1a_2-a_1}\right]\, . 
\label{B11}
\end{eqnarray} 

\bigskip 
\section{Heat noise integrals} 
\medskip

The integrals (\ref{hn2},\ref{hn3}) can be computed explicitly for $\beta_1=\beta_2\equiv \beta$.  
Using the variables (\ref{B1}) one finds  
\begin{eqnarray} 
A^Q_{\rm f} &=&  \frac{1}{2\pi \beta^3} \left [ C(a_1;\mu_1) + C(a_2;\mu_1)\right ]\, ,  
\label{C1}\\
B^Q_{\rm f} &=& 
\frac{1}{2\pi \beta^3} \frac{(a_1+a_2)}{(a_1-a_2)} 
\left [D(a_1;\mu_1) -D(a_2;\mu_1)\right ]\, , 
\label{C2}
\end{eqnarray} 
where 
\begin{eqnarray} 
C(a;\mu_1) &=& \frac{(\beta \mu_1)^2 }{a+1} + 2 \beta \mu_1 \ln \left [\frac{a}{a+1}\right ] - 2\li_2\left [-\frac{1}{a}\right ]\, , 
\label{C3}\\
D(a;\mu_1) &=& (\beta \mu_1)^2 \ln\left [\frac{a}{a+1}\right] 
-2\beta \mu_1 \li_2\left [-\frac{1}{a}\right ] + 2\li_3\left [-\frac{1}{a}\right ]\, . 
\label{C4}
\end{eqnarray} 
Substituting these expressions in (\ref{hn1}) and restoring the variables $\mu_\pm$ one gets 
\begin{eqnarray} 
\hnf(\beta, \mum,\mup,\tau) = \frac{\tau^2}{2\pi \beta^3} 
\left [C\left (\e^{-\beta(\mup +\mum)};\mup+\mum\right ) +  C\left (\e^{-\beta(\mup -\mum)};\mup+\mum\right )\right ] + 
\nonumber \\
\frac{\tau (1-\tau)}{2\pi \beta^3} \left [
\frac{\e^{-\beta(\mup +\mum)}+\e^{-\beta(\mup -\mum)}}{\e^{-\beta(\mup +\mum)}-\e^{-\beta(\mup -\mum)}}\right ]
\left [D\left (\e^{-\beta(\mup +\mum)};\mup+\mum\right ) +  D\left (\e^{-\beta(\mup -\mum)};\mup+\mum\right )\right ]\, . 
\nonumber \\
\label{C5}
\end{eqnarray}

\end{document}